\documentclass[a4paper,11pt]{article}
\usepackage{pos}
\usepackage{upgreek}
\usepackage{bm}
\usepackage{subcaption}
\usepackage{hyperref}
\usepackage{wrapfig}

\title{Lorentz invariance violation in the extragalactic propagation of ultra-high energy photons and in the development of showers in the Earth crust and atmosphere}
 \ShortTitle{LIV in UHE photons}

\author*[a,b]{D. Boncioli}
\author[c]{V. B. Bezerra}
\author[a,b]{M. Giammarco}
\author[c]{I. P. Lobo}
\author[c]{P. Morais}
\author[a,b]{F. Salamida}

\affiliation[a]{Università degli Studi dell’Aquila, Dipartimento di Scienze fisiche e chimiche, via Vetoio,
L’Aquila, Italy}
\affiliation[b]{Istituto Nazionale di Fisica Nucleare, Laboratori Nazionali del Gran Sasso, Assergi (AQ), Italy}
\affiliation[c]{Physics Department, Federal University of Paraíba, Caixa Postal 5008, 58059-900, João Pessoa, PB, Brazil.}

\emailAdd{denise.boncioli@univaq.it}

\abstract{We investigate the effects of Lorentz invariance violation (LIV) on photon interactions, considering both intergalactic propagation (Breit-Wheeler process) and atmospheric interactions (Bethe-Heitler process). By incorporating LIV into the theoretical framework, we analyze how it modifies key quantities such as the cross section, threshold energy, and mean free path of photons traveling through intergalactic space. In addition, we study its impact on extensive air showers initiated by high-energy photons, demonstrating that LIV can alter the cross section of the primary interaction in the atmosphere. Additionally, we also test the photon interactions in the Earth crust, to evaluate if they can induce upward-going showers. Our results highlight the necessity of accounting for both propagation effects in intergalactic space and interactions in the atmosphere when evaluating LIV signatures. Even small deviations from Lorentz invariance can lead to measurable changes in astroparticle propagation and photon dynamics, offering new opportunities to test quantum gravity theories through high-energy astrophysical observations.}

\ConferenceLogo{PoS_ICRC2025_logo}

\FullConference{%
39th International Cosmic Ray Conference (ICRC2025)\\
15 -- 24 July, 2025\\
Geneva, Switzerland}

\begin{document}
\maketitle

\section{Introduction}\label{intro}

Exploring quantum gravity effects remains a major challenge, as these theories are difficult to be tested experimentally, due to limitations in accessing the quantum gravity scale (see \cite{Addazi:2021xuf} and references therein for an overview). To address these issues, quantum gravity phenomenology seeks to parametrize departures from fundamental aspects of special and general relativity from the semiclassical limit of quantum gravity. Among possible deviations, Lorentz invariance violation (LIV) represents an important window of investigation since Planck scale effects can be amplified when analyzing high-energy particle interactions \cite{Coleman:1998ti}.

High-energy multimessenger scenarios, including cosmic rays, gamma rays, neutrinos, and gravitational waves, provide the most effective tools for testing fundamental physics due to their high energy in the laboratory frame \cite{Aloisio:2000cm,Coleman:1998ti}. In this context, modifications due to LIV can influence various processes relevant to the production, propagation, and detection of these messengers. In these scenarios, modified dispersion relations (MDRs) frequently arise as a consequence of LIV, and can be expressed in terms of energy as follows\footnote{Natural units are used hereinafter.}:
\begin{equation}
    E^2 = m^2 +p^2+m_{\mathrm{eff}}^2,\quad\quad m_{\mathrm{eff}}^2=\sum_{n\geq0} \eta_{i,n}\frac{E^{n+2}}{M_{\mathrm{Pl}}^n}\label{mdr}
\end{equation}
where $m_{\mathrm{eff}}$ is the effective momentum-dependent mass of the particle, $\eta_{i,n}$ is the LIV parameter (corresponding to the type of particle $i$ and order $n$), $M_{\text{Pl}} = 1.22 \times 10^{19}\,\text{GeV}$ is the Planck mass.

In this contribution, we explore the propagation of photons in the extragalactic space and the Earth's atmosphere, by focusing separately on the modifications in the energy threshold for the relevant reactions as well as in the cross sections, in order to improve the analyses for testing LIV in UHE photons as also investigated in \cite{Martinez-Huerta:2020cut,PierreAuger:2021tog,Lang:2022rgb,Carmona:2024thn}. A preliminary version of these results was already shown in \cite{Morais:2025zre}. We also investigate, for the first time, the possibility that LIV photons initiate a shower in the Earth crust and contribute to the upward-going showers as neutrino-like particles.

\section{The effects of LIV in the extragalactic propagation}\label{LIVextragalprop}

Photons produced in extragalactic space as a result of interactions between cosmic rays and the cosmic microwave background (CMB) or the extragalactic background light (EBL) are typically high-energy photons with $E_\gamma = [10^9, 10^{22}]$ eV. They can interact again with background photons -- specifically those of the CMB and the radio background (RB), which have much lower energies, in the range $\epsilon = [10^{-11}, 10]$ eV -- via the pair-production process $\gamma\gamma_{CMB} \rightarrow e^+e^-$. Here $\gamma$ represents the propagating photon with energy $E_\gamma$, while $\gamma_{CMB}$ corresponds to the lower-energy CMB photon with energy $\epsilon$, producing electron-positron pairs and initiating an electromagnetic cascade.

Considering the cosmological effects as negligible, the optical depth, $\tau(E_\gamma,z_s)$, of the pair-production process can be written as:
\begin{equation}
    \tau_\gamma(E_\gamma,z_s) = \int_0^{z_s}  dz \frac{dl(z)}{dz}\int^1_{-1}d(\cos{\theta})\frac{1-\cos{\theta}}{2}\int_{\epsilon_{th}}^{\infty}\sigma(E_\gamma,\epsilon)n_\gamma(\epsilon,z) d\epsilon,\label{optical_depth}
\end{equation}
where $\theta$ is the angle between the direction of the propagating photon and the one of the background ($\theta = [-\pi,+\pi]$), $\sigma(E_\gamma,\epsilon)$ is the cross section of the interaction, $\epsilon_{th}$ is the threshold energy\footnote{The distance traveled by a photon per unit redshift, at redshift $z$, is 
\begin{equation}
    \frac{dl(z)}{dz} = \frac{c}{H_0(1+z)\sqrt{\Omega_\Lambda+\Omega_M(1+z)^3}},
\end{equation}
where $\Omega_\Lambda = 0.7$ is the dark energy density, $\Omega_M = 0.3$ is the matter density, $H_0 = 70\,\text{km}\,\text{s}^{-1}\,\text{Mpc}^{-1}$ is the Hubble constant and $c$ is the speed of light in vacuum.}. LIV can manifest itself both in the threshold energy of the process as well as in the cross section, as it will be discussed in the next sections.

As suggested in \cite{DeAngelis:2013jna}, we compute for a photon $\gamma$ with energy $E_\gamma$ and at a given redshift, $z_s$, the probability to reach Earth without interacting with the background as:
\begin{equation}
    P_{\gamma\rightarrow\gamma}(E_\gamma,z_s) = e^{-\tau_{\gamma}(E_\gamma,z_s)}.
\end{equation}

\subsection{LIV modifications at the threshold energy}
The dispersion relation for the photons, when LIV is included, can be expressed as: $E_\gamma^2 - p_\gamma^2 = m_{\gamma, \mathrm{eff}}^2$, 
where $E_\gamma$ is the energy, $p_\gamma$ is the momentum, and $m_{\gamma,\mathrm{eff}}$ is the effective mass of the photon. For background photons, the energy is given by $\epsilon = p_{\gamma_{CMB}}$. Since their energy is significantly smaller than the Planck energy, the effects on their dispersion relations can be neglected. For electron-positron pairs, the dispersion relation follows the standard form: $E^2_{e^\pm}-p^2_{e^\pm} = m^2_{e^\pm}$.

To analyze how LIV impacts the interaction, we consider the conservation of energy and momentum, leading to the expression of the squared energy, $s$.
\begin{equation}
s = 2E_\gamma\epsilon\left(1-\cos{\theta}\right) + m_{\gamma, \mathrm{eff}}^2,\label{squared_s}
\end{equation}
where the minimum energy required for the pair production is modified. For a head-on collision ($\theta = \pi$) in Eq.~\ref{squared_s}, the threshold condition becomes:
\begin{equation}
\epsilon_{\text{th}} \geq \frac{4m_e^2 - m_{\gamma, \mathrm{eff}}^2}{4E_\gamma}.\label{Eq.threshold_energy}
\end{equation}

\begin{figure}[!t]
\centering
    \begin{subfigure}[b]{0.48\textwidth}
    \centering
    \includegraphics[width=7.8cm]{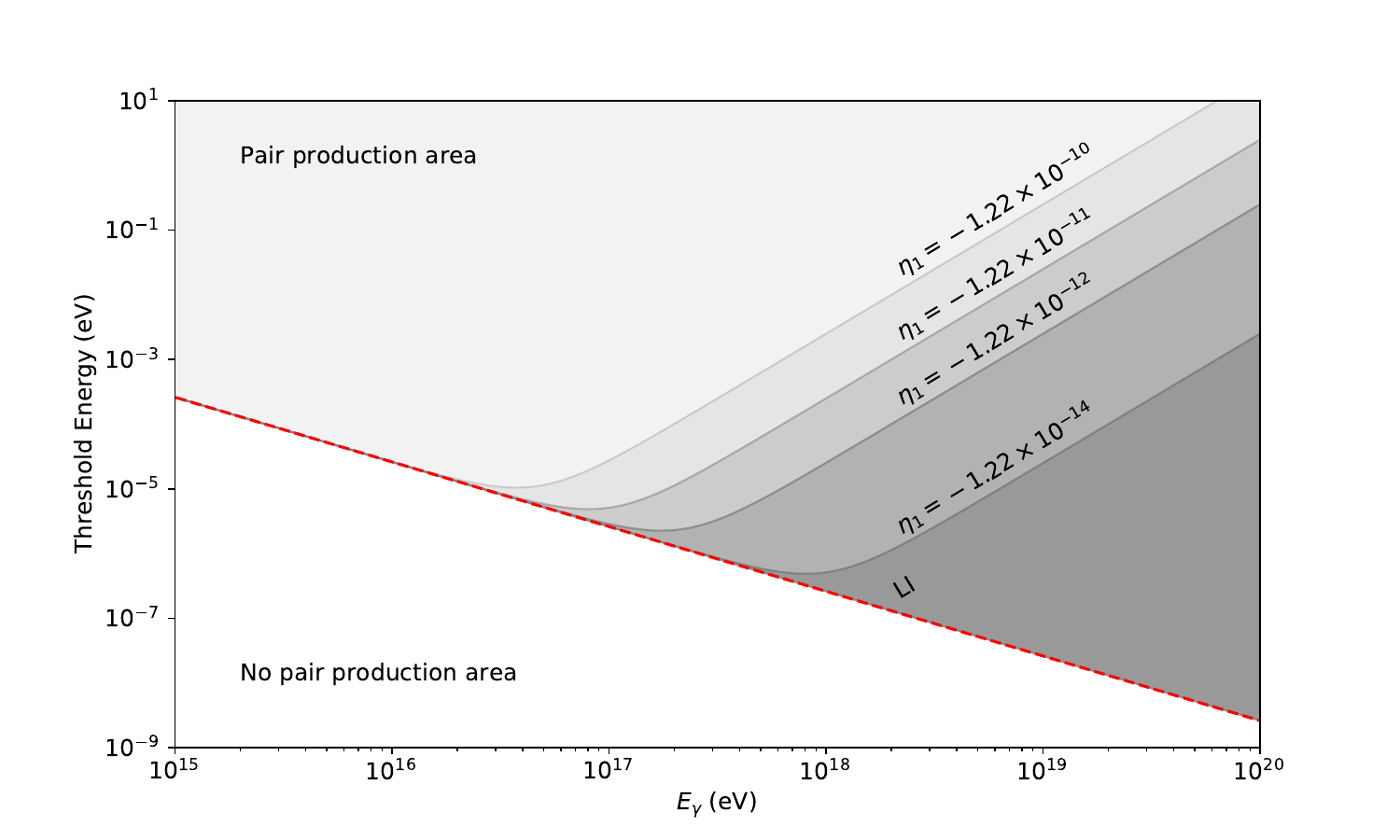}
    \caption{}\label{threshold}
    \end{subfigure}
    \quad
    \begin{subfigure}[b]{0.48\textwidth}
    \centering
    \includegraphics[width=7cm]{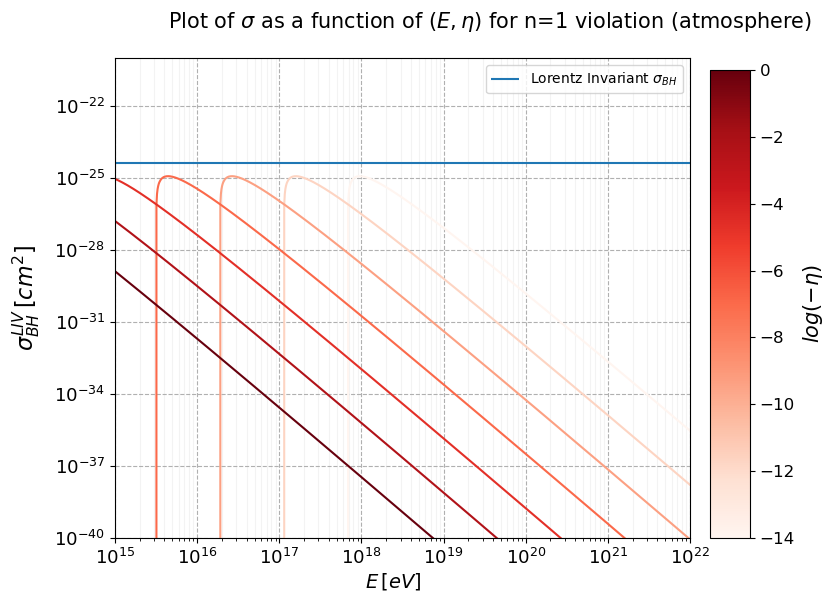}
    \caption{}\label{xsection}
    \end{subfigure}
    \caption{(a) Threshold energy for pair production as a function of the energy of the high-energy photon $E_\gamma$, showing allowed regions for different $\eta$ values. Darker to lighter gray shades represent increasing $\eta$ for $n=1$, with the red dashed line for $\eta = 0$ (LI case). (b) Cross section for pair production in the atmosphere as a function of the energy of the high-energy photon $E_\gamma$, for different $\eta$ values.}
\label{th-xsect}
\end{figure}

The variation of the threshold energy for different values of $\eta$ (which enters in $m_{\gamma, \mathrm{eff}}$ for $n=1$) is illustrated in Fig.~\ref{threshold}, demonstrating that the region where the process is allowed increases as $\eta$ decreases.

\subsection{LIV modifications at the cross section}
The cross section for the process $\gamma\gamma_{CMB}\rightarrow e^+e^-$ is described by the Breit-Wheeler process and, in the standard case, can be written as:
\begin{equation}
    \sigma_{BW}(E_\gamma,\epsilon) = \frac{\alpha^2\pi}{2m_e^2}(1-\beta^2)\left[2\beta(\beta^2-2)+(3-\beta^4)\log\left(\frac{1+\beta}{1-\beta}\right)\right],\label{sigma_st}
\end{equation}
with $\alpha = 1/137$, and $\beta$ is the velocity of the electron and positron in the center of mass reference frame $\beta(E_\gamma,\epsilon) = \sqrt{1-\frac{4m_e^2}{s}}$.

Now, considering the aspects of LIV, we modify the cross section as described in \cite{Rubtsov:2012kb,Rubtsov:2013wwa}: 
\begin{equation}
    \sigma_{BW}^{LIV} = \frac{\alpha^2 \pi}{E_\gamma \epsilon (1-\cos\theta)} \left(1+\left(1+\frac{2 m_{\gamma, \mathrm{eff}}^2}{E_\gamma \epsilon (1-\cos\theta)}\right)^2\right)\log\left(\frac{E_\gamma \epsilon (1-\cos\theta)+m_{\gamma \mathrm{eff}}^2}{m_e^2}\right);\label{sigma_liv}
\end{equation}
this expression is valid as long as $\eta \gg  (m_e^2-2E_\gamma \epsilon)\frac{M_{Pl}^n}{E_\gamma^{n+2}}$. Fig.~\ref{xsection} shows how the cross section changes as a function of the energy, for several $\eta$ values.

\subsection{The modified mean free path in the extragalactic space} 
Following the approach of \cite{DeAngelis:2013jna}, the survival probability for photons of energy $E$ from a source at a redshift $z$ can be approximated as $P_{\gamma\rightarrow\gamma}(E_\gamma,D)\approx \exp(-D/\lambda_{\gamma}(E))$, where $D = cz_s/H_0$ is the distance from the source  and the mean free path for the VHE photons is then given by
\begin{equation}
\frac{1}{\lambda(E_\gamma)} = \frac{\tau_\gamma(E_\gamma, DH_0/c)}{D}\label{mfp}.
\end{equation}
The inverse of the mean free path can be expressed in a simple form,
\begin{equation}
    \frac{1}{\lambda_{\gamma}} = \frac{2 (kT)^3}{\pi^2\Bar{E}^2}\int_1^\infty\left(\Bar{s}-\frac{m_{\gamma\,eff}^2}{4m_e^2}\right)\,d\Bar{s}\int_{\Bar{\epsilon}_{th}}^\infty \frac{d\Bar{\epsilon}}{e^{\Bar{\epsilon}}-1}\,\sigma,\label{mfp_simple}
\end{equation}
where the product of the Boltzmann constant $k$ and the temperature $T$ of the CMB is $kT = 2.35\, \times \, 10^{-4}$ eV, and $m_e^2 = 2.61\times10^{11}\,\text{eV}^2$.
For convenience we have used the dimensionless variables 
\begin{equation}
    \Bar{s}=\frac{s}{4m_e^2},\,\,\,\Bar{\epsilon}=\frac{\epsilon}{kT},\,\,\,\Bar{E}=\frac{E_\gamma}{m_e^2/(kT)},\label{change_variables}
\end{equation}
so that the new limits of integration in terms of the dimensionless variables are  $1 \leq \Bar{s}\leq \infty$ and $(\Bar{s}-\frac{m_{\gamma, \mathrm{eff}}^2}{4m_e^2})\Bar{E}^{-1} = \Bar{\epsilon}_{th} \leq \Bar{\epsilon} \leq \infty$.

\begin{figure}[!t]                     
\centering
\includegraphics[width=0.7\textwidth]{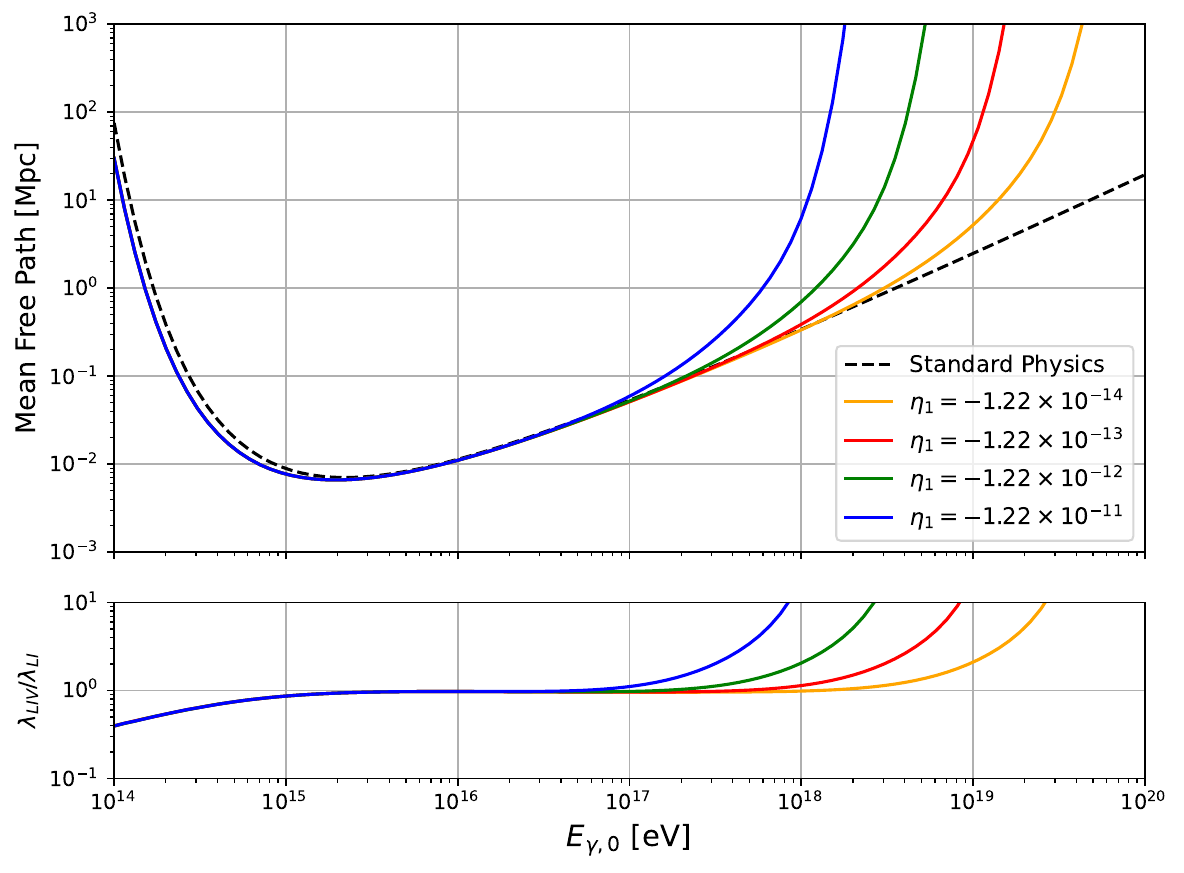}
\begin{minipage}[t]{.24\textwidth}
\vspace{-6.6cm} 
\caption{Mean free path of photons as a function of their energy in the range where the CMB is the relevant background. The black curve corresponds to the SR case, and the colored curves represent the LIV scenarios with modifications in cross section and threshold for different values of $\eta$ and $n=1$.}
\label{MFP_liv}
\end{minipage}
\end{figure}

At this point, the LIV effects in the mean free path are in the threshold energy. Now, the cross section in Eq.~\ref{mfp_simple} can be written considering the expression from Eq.~\ref{sigma_liv}, where we can assume a head-on collision, and in terms of the dimensionless variables of Eq.~\ref{change_variables}. The results of the LIV modifications in the mean free path are shown in Fig.~(\ref{MFP_liv}). 
Notably, the mean free path increases significantly for high-energy photons as $\eta$ increases, demonstrating the suppression of photon interactions at ultra-high energies. 

\section{The effect of LIV in the development of showers in the Earth atmosphere and crust}

Primary ultra-high-energy (UHE) photons arriving at Earth interact with the atmosphere, initiating extensive air showers. For photon energies around $10^{18}$ and $10^{19}$ eV, the predominant interaction mechanism is the production of $e^+e^-$ pairs on nitrogen nuclei in the atmosphere, known as the Bethe-Heitler (BH) process. Under standard conditions, the cross section for this process can be described as energy-independent as follows:
\begin{equation}
    \sigma_{BH} = \frac{28 Z^2\alpha^3}{9m_e^2}\left(\log\frac{183}{Z^{\frac{1}{3}}}-\frac{1}{42}\right),
    \label{sigmaBH}
\end{equation}
where $m_e$ is the electron mass, $\alpha = 1/137$ is the fine-structure constant, and $Z$ is the atomic number of the target nucleus. Considering nitrogen $(Z = 7)$ as the target, the cross section is $\sigma_{BH} \approx 500$ mb. Now, if we account for the effects of LIV with the MDR in Eq.~\ref{mdr}, the modified cross section studied in \cite{Rubtsov:2012kb,Rubtsov:2013wwa} takes the form:
\begin{equation}
    \sigma_{BH}^{LIV} = \frac{8 Z^2\alpha^3}{3|m_{\gamma, \mathrm{eff}}^2|}\log\frac{1}{\alpha Z^{\frac{1}{3}}}\log\frac{|m_{\gamma, \mathrm{eff}}^2|}{m_e^2},
\end{equation}
where the $|m_{\gamma, \mathrm{eff}}^2|$ is the effective momentum-dependent mass. The validity of this modification is determined by the condition $|m_{\gamma, \mathrm{eff}}^2|\gg m^2_e$ according to \cite{Rubtsov:2013wwa}, which results in $|\eta|\gg m_e^2 \frac{M_{Pl}^n}{E^{n+2}}$.
The probability for a photon to produce a pair in the atmosphere, as shown in \cite{Satunin:2023yvj}, can be written as,
\begin{equation}
    P = \int^{X_{\text{atm}}}_0 dX_0\frac{e^{- X_0/ \langle X_0 \rangle_{\text{LIV}}}}{\langle X_0 \rangle_{\text{LIV}}} = 1-e^{-X_\text{atm}/\langle X_0 \rangle _{\text{LIV}}},\label{patm}
\end{equation}
where $X_{\mathrm{atm}}$ is the overall atmospheric depth, and the mean depth of interactions $\langle X_0\rangle _\text{LIV}$ in the LIV case can be expressed via the mean depth $\langle X_0 \rangle_\text{LI} = 57\,\text{g}\,\text{cm}^{-2}$ in the LI case and the ratio of the cross sections $\sigma^\text{LIV}$ and $\sigma^\text{LI}$ as follows:
\begin{equation}
    \langle X_0\rangle_\text{LIV} = \frac{\sigma^\text{LI}}{\sigma^\text{LIV}}\langle X_0 \rangle_\text{LI}.\label{crossection}
\end{equation}

As well as what happens in the atmosphere, UHE photons are in principle able to initiate a cascade of secondary particles in the Earth crust, through the Bethe-Heitler process whose cross section is reported in Eq.~\ref{sigmaBH}, the only difference being the average charge of the involved nuclei. We aim therefore to estimate the probability for a UHE photon to cross the Earth, as a function of the incident angle, with a similar approach as the one used in \cite{PierreAuger:2023wke} for neutrinos. The reason for this is that, if photons survive within the Earth crust, and interact near the Earth surface, they could be responsible of upward-going showers. We express the probability as:
\begin{equation}
P_{\mathrm{surv,Earth}}(\theta)=\exp\left(- \frac{N_A \sigma^{\mathrm{LIV}}_{\mathrm{BH}}}{M} \sum_{i=1}^{n(\theta)} l_{i}(\theta) \rho_i\right),
\label{PEarth_surv}
\end{equation}
where $n(\theta)$ is the number of crossed layers in the PREM model\footnote{PREM (Preliminary Reference Earth Model) \cite{Dziewonski:1981xy} models the Earth density profile with slab approximations, assuming constant density in each layer.}, $l_i(\theta)$ is the propagation length in the layer $i$ and $M$ is the average molar mass, taken as $\approx 26.5\,\mathrm{g\,mol^{-1}}$. The probability of interacting in the last $d=50$ km of crust is instead
\begin{equation}
P_{\mathrm{int,Earth}}(\theta)=1-\exp\left(- \frac{N_A \sigma^{\mathrm{LIV}}_{\mathrm{BH}}}{M} \sum_{i=1}^{n(\theta)} d_{i}(\theta) \rho_i\right).
\label{PEarth_int}
\end{equation}
The total probability for the UHE photon to propagate in the Earth and interact in the last 50 km is therefore:
\begin{equation}
P_{\mathrm{gen}}(\theta)=P_{\mathrm{int,Earth}}(\theta)P_{\mathrm{surv,Earth}}(\theta),
\label{PEarth_gen}
\end{equation}
and is shown in Fig.~\ref{Pgen}. The non-observation of upward-going showers by the Auger Observatory could therefore be used to constrain LIV effects which can in principle enhance the probability of initiating a shower within the Earth crust in addition to the only atmosphere.

\section{Results and discussion}
\begin{figure}[!t]
\centering
    \begin{subfigure}[b]{0.48\textwidth}
    \centering
    \includegraphics[width=7.5cm]{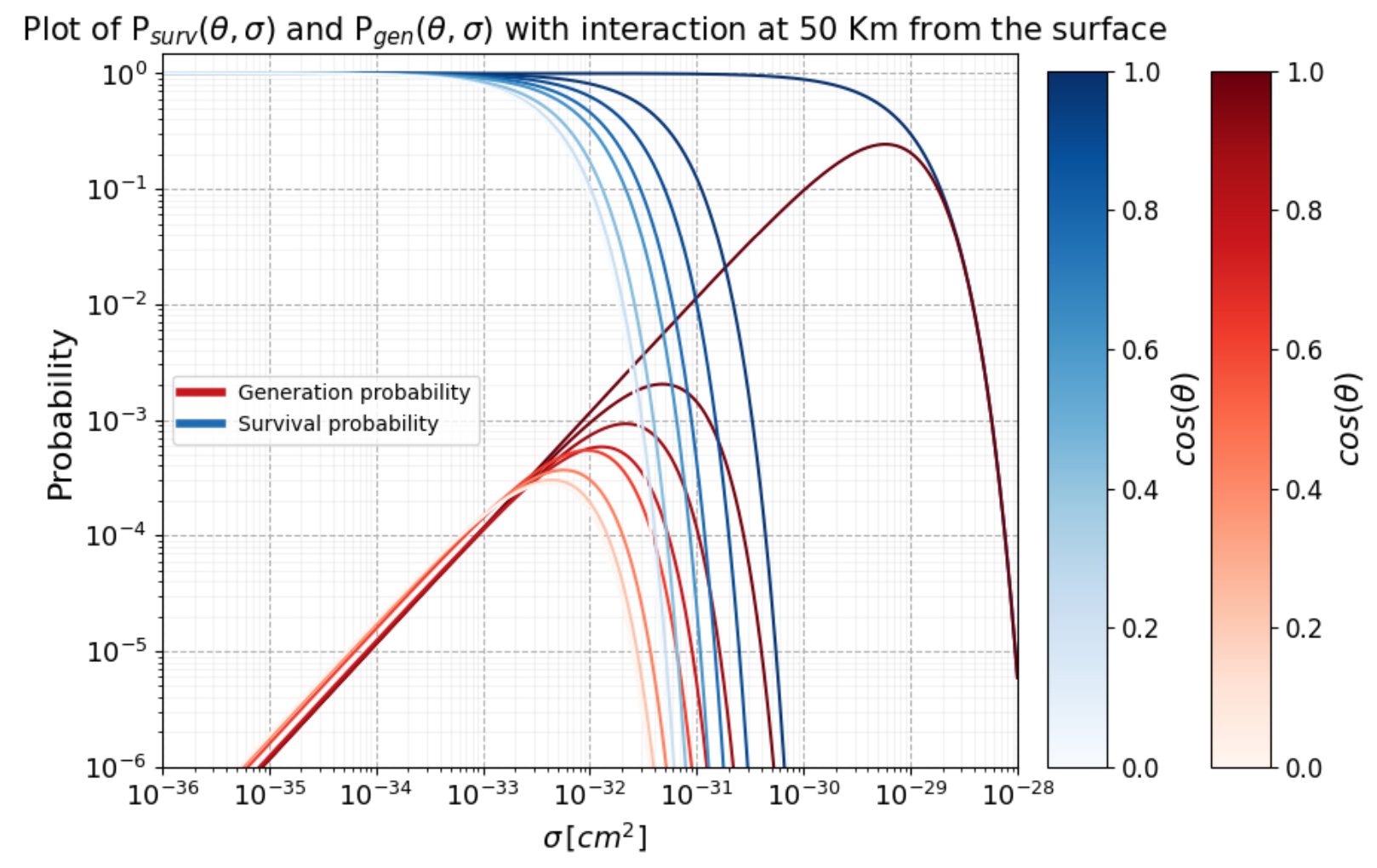}
    \caption{}\label{Pgen}
    \end{subfigure}
    \quad
    \begin{subfigure}[b]{0.48\textwidth}
    \centering
    \includegraphics[width=6.8cm]{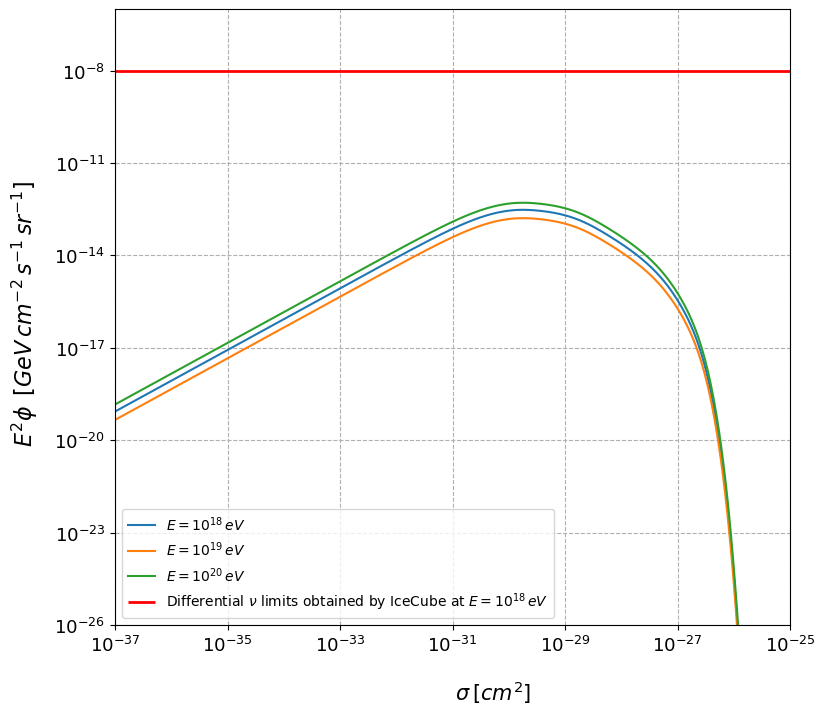}
    \caption{}\label{flux_sigma}
    \end{subfigure}
    \caption{(a) Survival probability (blue) for a photon in the Earth crust, and generation probability (red), as a function of the BH modified cross section, for different incident angles (as from Eq.~\ref{PEarth_int},~\ref{PEarth_surv},~\ref{PEarth_gen}); (b) Differential UHE photon flux as a function of the BH cross section, for the case of interactions in the Earth crust, for three values of energy of the UHE photon.}
\end{figure}

As formulated in \cite{Satunin:2023yvj}, in order to estimate the effective observable flux with LIV effects, we compute the following modified differential photon flux as:
\begin{equation}
    \left(\frac{d\Phi(E,\eta)}{dE}\right)_{\text{LIV,i}} = P(E,\eta)_i \frac{d\Phi(E,\eta)}{dE}\bigg{|}_{\text{top-of-atm}},
\end{equation}
where the suppression factor is given by Eq.~\ref{patm} in case of the LIV effect in the atmosphere or by Eq.~\ref{PEarth_gen} if the effect in the Earth is considered (in the latter, the probability is integrated over the angles of incidence, for values $\theta=[0.23^{\circ},5^{\circ}]$ in order to account for at least 50 km of path within the Earth crust). The flux of photons at the top of the atmosphere is in any case dependent on LIV effects, as the ones related to the extragalactic propagation, described in Sec.~\ref{LIVextragalprop}. 

Taking into account the results obtained in \cite{PierreAuger:2021tog} for the photon flux with several LIV parameters, we compute the expected fluxes including LIV in the Earth crust, as a function of the modified value of the cross section; Fig.~\ref{flux_sigma} shows the result for three values of energy, compared to the experimental differential upper limits for neutrinos at energy $=10^{18}$ eV set by IceCube. For a fixed energy, corresponding to an increase of the LIV parameter, the cross section decreases, therefore the interactions are less probable and the expected flux decreases. Although the outcome of this preliminary study shows that the expected flux is quite below the IceCube limit, the method is worth to be better explored as a new way to investigate LIV in the detection stage of astroparticles.

\begin{figure}[!t]
\centering
    \begin{subfigure}[b]{0.48\textwidth}
    \centering
    \includegraphics[width=7cm]{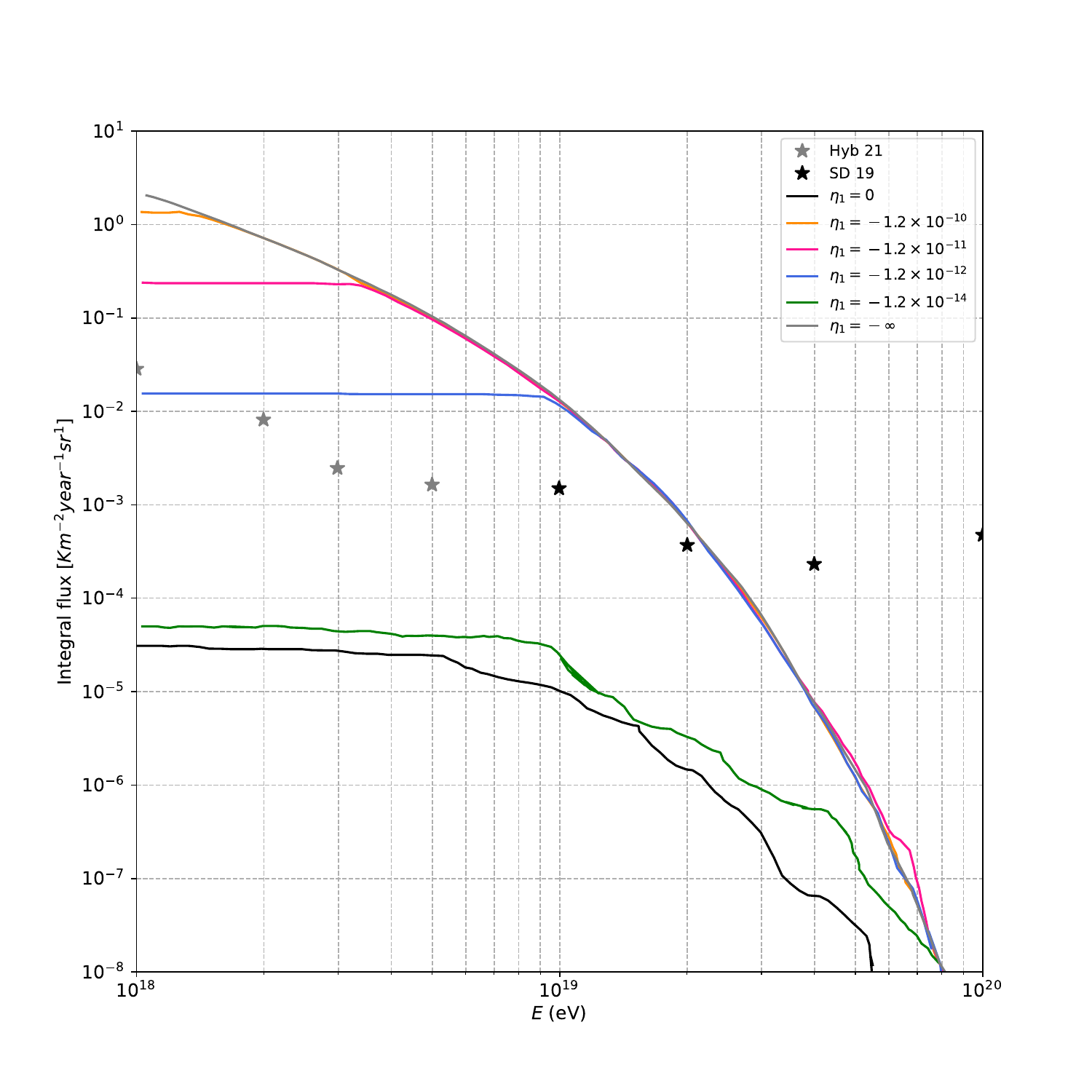}
    \caption{}\label{AugerLIVpaper}
    \end{subfigure}
    \quad
    \begin{subfigure}[b]{0.48\textwidth}
    \centering
    \includegraphics[width=7cm]{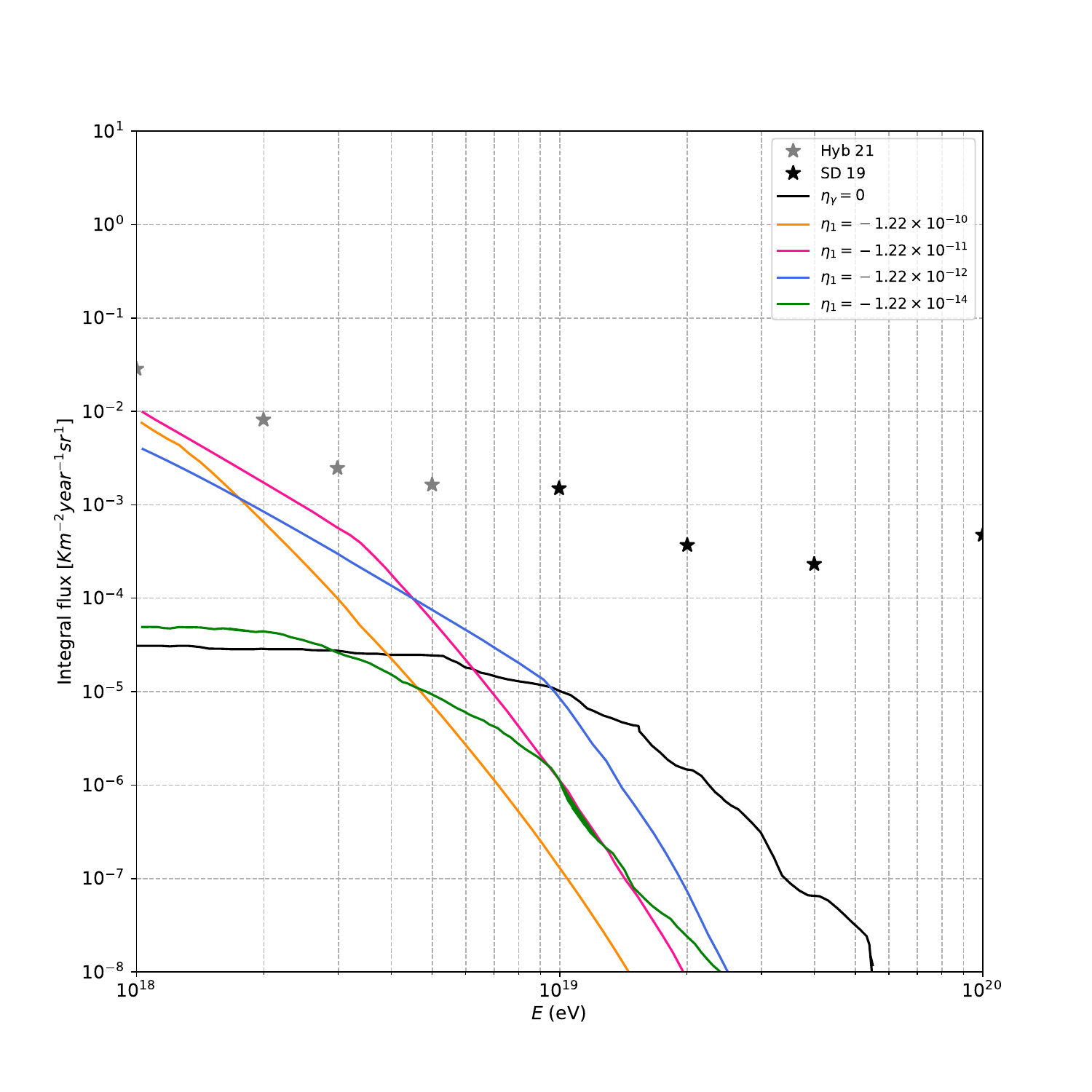}
    \caption{}\label{AugerLIVpaper-atm}
    \end{subfigure}
    \caption{Integral flux of UHE photons as a function of the energy, for several LIV values, compared to the upper limits from the Pierre Auger Observatory \cite{Savina:2021cva}, (a) as computed in \cite{PierreAuger:2021tog}, with LIV effects only in the extragalactic propagation, and (b) including the atmospheric suppression as calculated in this work.}
\label{int_flux_prop_atmo}
\end{figure}

While the interactions of photons in the Earth crust are expected to contribute to the upward-going showers and as a consequence must be compared to expectations for neutrinos, the LIV modified interactions in the atmosphere alter the expected photon flux. The effect of the atmospheric suppression on the simulated UHE photon flux is shown in Fig.~(\ref{AugerLIVpaper-atm}), in contrast to Fig.~(\ref{AugerLIVpaper}), which does not account for atmospheric suppression. The results show that including the suppression of atmospheric shower production reduces the expected photon flux, aligning the curves with the measured upper limits from the Pierre Auger Observatory for the studied values of  $\eta$. This suggests that the parameter space which can be excluded if the whole extragalactic and atmospheric effects are accounted for is smaller than what found in \cite{PierreAuger:2021tog}.

To conclude, in this work we explore the effects of LIV on the propagation of UHE photons in extragalactic space and their interaction in Earth atmosphere and crust. We analyze how LIV modifies both the cross-section and threshold energy for the BW and BH processes. Although LIV increases the number of UHE photons reaching Earth, it also suppresses the formation of particle showers in the atmosphere. The constraints on LIV parameters become less strict when accounting for atmospheric suppression, however the results are more robust as they include LIV effects in both propagation and detection stages.

\end{document}